\documentclass[aps,pra,reprint]{revtex4-2}
\usepackage{amsmath}
\usepackage{amssymb}
\usepackage{mathrsfs}
\usepackage{graphicx}
\usepackage{booktabs}  
\usepackage{multirow}  
\usepackage[colorlinks=true, letterpaper=true, pdfstartview=FitV, linkcolor=blue, citecolor=blue, urlcolor=blue]{hyperref}

\begin{document}

\title{Broadband optical nonreciprocity via nonreciprocal band structure}
\author{Ning Hu}
\affiliation{Key Laboratory of Low-Dimensional Quantum Structures and
Quantum Control of Ministry of Education, Key Laboratory for Matter
Microstructure and Function of Hunan Province, Department of Physics and
Synergetic Innovation Center for Quantum Effects and Applications, Hunan
Normal University, Changsha 410081, China}
\author{Zhi-Xiang Tang}
\affiliation{Key Laboratory of Low-Dimensional Quantum Structures and
Quantum Control of Ministry of Education, Key Laboratory for Matter
Microstructure and Function of Hunan Province, Department of Physics and
Synergetic Innovation Center for Quantum Effects and Applications, Hunan
Normal University, Changsha 410081, China}
\author{Xun-Wei Xu}
\email{xwxu@hunnu.edu.cn}
\affiliation{Key Laboratory of Low-Dimensional Quantum Structures and
Quantum Control of Ministry of Education, Key Laboratory for Matter
Microstructure and Function of Hunan Province, Department of Physics and
Synergetic Innovation Center for Quantum Effects and Applications, Hunan
Normal University, Changsha 410081, China}

\date{\today}

\begin{abstract}
As a promising approach for optical nonreciprocity without magnetic materials, optomechanically induced nonreciprocity has great potential for all-optical controllable isolators and circulators on chips.
However, as a very important issue in practical applications, the bandwidth for nonreciprocal transmission with high isolation has not been fully investigated yet.
In this study we review the nonreciprocity in a Brillouin optomechanical system with single cavity and point out the challenge in achieving broad bandwidth with high isolation.
To overcome this challenge, we propose a one dimensional optomechanical array to realize broadband optical nonreciprocity via nonreciprocal band structure.
We exploit nonreciprocal band structure by the stimulated Brillouin scattering induced transparency with directional optical pumping, and show that it is possible to demonstrate optical nonreciprocity with both broad bandwidth and high isolation.
Such Brillouin optomechanical lattices with nonreciprocal band structure, offer an avenue to explore nonreciprocal collective effects in different electromagnetic and mechanical
frequency regimes, such as nonreciprocal topological photonic and phononic phases.
\end{abstract}

\maketitle

\section{Introduction}

Cavity optomechanics for the optical and mechanical modes coupled through radiation-pressure interaction (for review, see Ref.~\cite{Aspelmeyer2014RMP})
is a rapidly developing field and have wide applications ranging from gravitational wave detections~\cite{Abbott2016PRL} to modern quantum technologies~\cite{Barzanjeh2022NatPh}.
Recent studies~\cite{2017NatPh922V} indicate that optomechanical system is an elegant candidate for implementing optical nonreciprocity without magnetic materials.
Based on the optomechanical interactions, many nonreciprocal devices, such as isolators and circulators, are proposed theoretically via various of mechanisms, including asymmetric optomechanical nonlinear interaction~\cite{Manipatruni2009PRL,2015NatSRWang,XuXW2018PRA,SongLN2019PRA}, directional enhanced optomechanical interaction in whispering-gallery-mode (WGM) microresonators~\cite{Hafezi12OE,LiBJ19PRJ,XuXW20PRJ,Jiao2020PRL,TangZX2023PRAPP}, synthetic magnetism for a closed loop of optical and mechanical modes with controllable phases~\cite{XuXW2015PRA,Metelmann2015PRX,Schmidt2015Optic,XuXW2016PRA,LiY2017OExpr,TianL2017PRA,JiangC2018PRA,Malz2018PRL,LiGL2018PRA,QianYB2021PRA,Seif2018NatCo,Barzanjeh2018PRL,Habraken2012NJPh,XuXW2020PRAPP,LaiDG2022PRL,LanYT2022OptL,LiuJX2023SCPMA}, and dynamical encircling of the exceptional point~\cite{ZhangJQ2022Nanop,LongD2022PRA}.
As a versatile platform, nonreciprocity has been realized in various optomechanical systems working in different frequency domains, ranging from optical regime with silica microsphere/microtoroid~\cite{Dong2015NatCo,Kim2015NatPh,Dong2016NaPho,Ruesink2016NatCo,Shen2018NatCo,Ruesink2018NatCo,ChenY2021PRL}, silicon nitride membrane
placed inside a high-finesse optical cavity~\cite{XuHT2016Natur,Doppler2016Natur,XuHT2019Natur}, and silicon optomechanical crystal circuit~\cite{Fang2017NatPh,Mathew2020NatNa,Pino2022Natur}, to the microwave regime implemented in superconducting microwave circuits~\cite{Peterson2017PRX,Bernier2017NatCo,Barzanjeh2017NatCo,Mercier2019PRAPP,Mercier2020PRL}.

As a essential parameter, the isolation of nonreciprocity has been seriously studied from both the theoretical and experimental aspects.
However, as another important parameter, the bandwidth of nonreciprocity has attracted much less attentions in the past studies~\cite{LiangC2020PRL}.
It has been shown that the bandwidth of the optomechanically induced nonreciprocity is ultimately limited by the optical linewidths~\cite{Hafezi12OE,Ruesink2016NatCo}.
So far, how to break the bandwidth limit of nonreciprocity is still an open question.
To answer this question, we propose to demonstrate optomechanical nonreciprocity in an optomechanical array with nonreciprocal band structure that simultaneously favors broad bandwidth and high reverse isolation.

In a very recent work~\cite{TangJS2022PRL}, nonreciprocal single-photon band structure was proposed in a one-dimensional (1D) coupled-resonator optical waveguide that chirally couples to an array of two-level quantum emitters. Inspired by this work~\cite{TangJS2022PRL} and the rapid growth of topological optomechanical lattices~\cite{RenHJ2022NatCo,Youssefi2021arXiv}, we propose to realize optomechanical nonreciprocity with high reverse isolation and bread bandwidth via nonreciprocal band structure in a Brillouin optomechanical array.
We note that optical nonreciprocity has been realized by the stimulated Brillouin scattering between the optical and
mechanical WGMs circulating along the equatorial surface with the control laser maintained in one direction~\cite{Dong2015NatCo,Kim2015NatPh}.
Here, we show that nonreciprocal band structure can be generated in a Brillouin optomechanical lattice with one unit cell consisting of a Brillouin optomechanical cavity coupling to a WGM microresonator,
which is obviously different from the nonreciprocal band structure based on chirally coupling to two-level quantum emitters~\cite{TangJS2022PRL}.
Our work is also different from the non-reciprocal phonon transport that was proposed in an array of optomechanical cavities
with time-reversal symmetry broken by the position-dependent phase (synthetic magnetic field)~\cite{Seif2018NatCo} or the interplay of photonic spin-orbit coupling~\cite{Sanavio2020PRB,Lemonde2019NJPh}.

This rest of the paper is organized as follows: In Sec.~\ref{BIN}, we give a brief review on the optical nonreciprocity based on stimulated Brillouin scattering in a WGM optomechanical system, and elaborate the challenge we face in achieving optical nonreciprocity with both broad bandwidth and high isolation in a single cavity.
We propose to realize nonreciprocal band structure in a Brillouin optomechanical array in Sec.~\ref{NBS}.
In Sec.~\ref{WBON}, we demonstrate the optomechanical nonreciprocity with high reverse isolation and broad bandwidth via the nonreciprocal band structure, show the advantageous of strong nonreciprocity via the nonreciprocal band structure to the nonreciprocity in the Brillouin optomechanical system with single cavity, and discuss the influences of backscattering effect on the nonreciprocity.
Finally, we summarize the results in Sec.~\ref{Con}.

\section{Photon-phonon-interaction induced nonreciprocity}\label{BIN}

\begin{figure*}[tbp]
\includegraphics[bb=18 163 580 725, width=15 cm, clip]{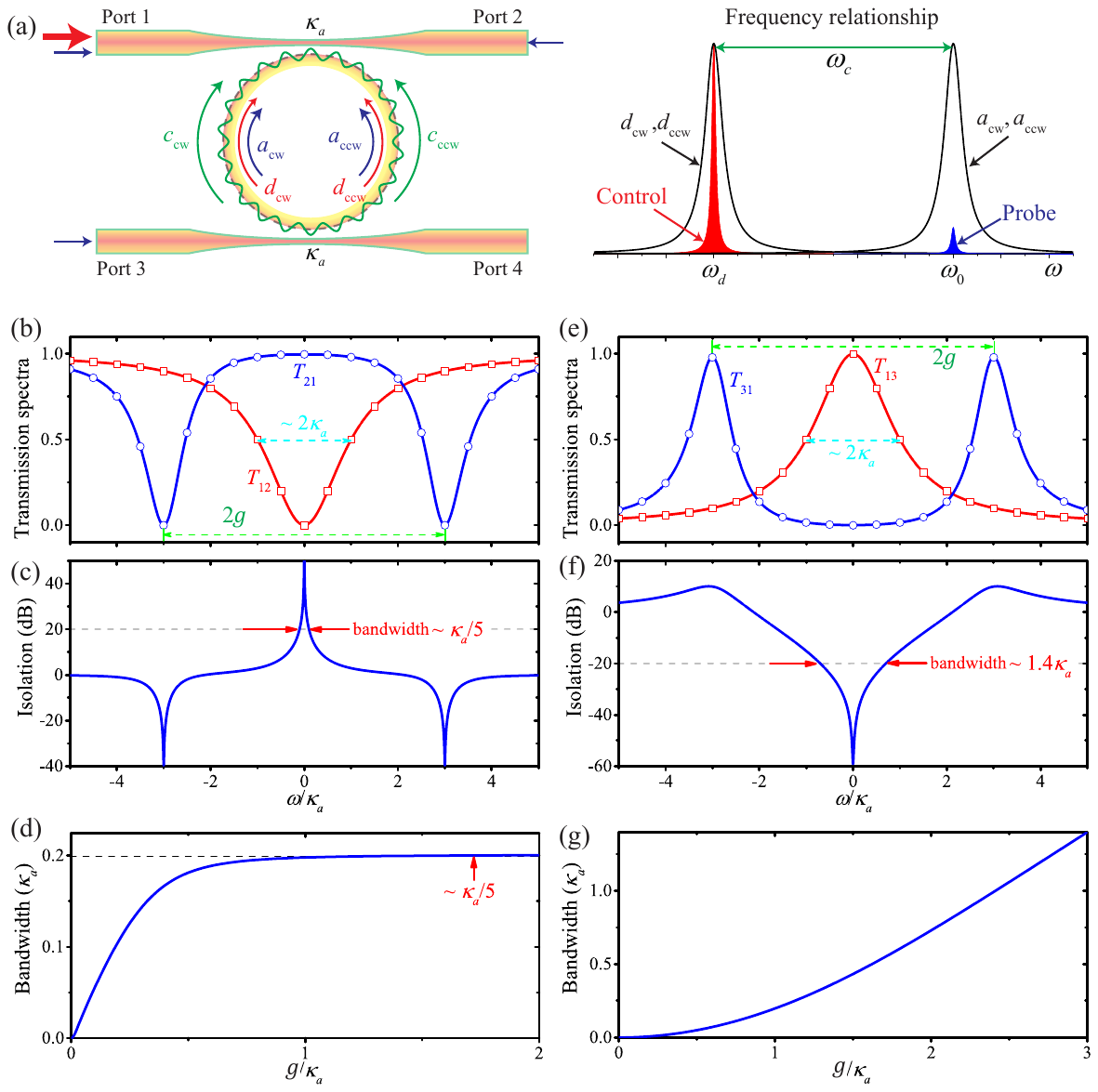}
\caption{(Color online) (a) Schematic illustration of the forward Brillouin optomechanical interaction and the frequency relationship under triple resonance $\omega_0=\omega_d+\omega_c$. (b) [(e)] The transmission
spectra ($T_{12}$ and $T_{21}$) [($T_{13}$ and $T_{31}$)] and (c) [(f)] the
isolation ($T_{21}/T_{12}$) [($T_{31}/T_{13}$)] for $g=3\protect\kappa_{a}$.
(d) [(g)] The bandwidth for the $20$ dB isolation ($T_{21}/T_{12}=100$) [($%
T_{31}/T_{13}=0.01$)] versus coupling $g/\protect\kappa_{a}$. Here, we set $%
\protect\kappa_{c}=\protect\kappa_{a}/100$.}
\label{fig1}
\end{figure*}

We start by reviewing optical nonreciprocity induced by photon-phonon
interactions in a microresonator supporting pairs of degenerate clockwise
(CW) and counter-clockwise (CCW) whispering-gallery optical modes, which has been
realized by several different experimental groups based on Brillouin
scattering~\cite{Dong2015NatCo,Kim2015NatPh} or optomechanical interaction~\cite{Dong2016NaPho,Ruesink2016NatCo,Shen2018NatCo,Ruesink2018NatCo}. We will discuss the
bandwidth for nonreciprocal transport between different ports with high isolation,
and show the challenge we face in achieving optical nonreciprocity with
both broad bandwidth and high isolation in a single cavity.

As a specific example, we consider a Brillouin optomechanical system that
consists of a microcavity supporting both optical and mechanical whispering
gallery modes (WGMs) travelling along the surface, and the optical modes are
evanescently coupled with two tapered fibres, as illustrated schematically
in Fig.~\ref{fig1}(a). The optomechanical system based on
forward Brillouin scattering can be described with the Hamiltonian
\begin{eqnarray}  \label{Eq1}
H_{\mathrm{Bom}} &=&\sum_{\sigma =\mathrm{cw,ccw}}\left( \omega
_{0}a_{\sigma }^{\dag }a_{\sigma }+\omega _{c}c_{\sigma }^{\dag }c_{\sigma
}+\omega _{d}d_{\sigma }^{\dag }d_{\sigma }\right)  \notag \\
&&+\sum_{\sigma =\mathrm{cw,ccw}}g_{b}\left( d_{\sigma }a_{\sigma }^{\dag
}c_{\sigma }+d_{\sigma }^{\dag }a_{\sigma }c_{\sigma }^{\dag }\right)  \notag
\\
&&+\left( \Omega e^{-i\omega _{p}t}d_{\mathrm{cw}}^{\dag }+\Omega ^{\ast
}e^{i\omega _{p}t}d_{\mathrm{cw}}\right) ,
\end{eqnarray}%
where $a_{\sigma }$ and $d_{\sigma }$ are two optical modes coupled through
Brillouin scattering mediated by the travelling acoustic wave $c_{\sigma }$
in the same direction. Here, $(\omega_{0},k_{0})$ and $(\omega_{d},k_{d})$
are the energies and momenta of the optical modes $a_{\sigma }$ and $%
d_{\sigma }$, and $(\omega_{c},k_{c})$ are the energy and momentum of the
travelling acoustic mode. To observe the forward Brillouin scattering
effect, the energies and momenta of these three modes must satisfy the
energy and momentum conservations $\omega_{c}=\omega_{0}-\omega_{d}$ and $%
k_{c}=k_{0}-k_{d}$ simultaneously. In order to enhance the single-photon
Brillouin coupling rate $g_b$, a strong control laser ($\Omega$ and $\omega
_{p}$) is input from Port 1. For simplicity, we assume that the
frequencies of the modes satisfy the resonant conditions $\omega _{p}=\omega
_{d}=\omega _{0}-\omega _{c}$ [Fig.~\ref{fig1}(a)].

In a rotating frame defined by the unitary transformation operator $%
R_1\left( t\right)= \exp (-iH_0 t)$ with $H_0=\sum_{\sigma =\mathrm{cw,ccw}%
}\left( \omega _{0}a_{\sigma }^{\dag }a_{\sigma }+\omega _{c}c_{\sigma
}^{\dag }c_{\sigma }+\omega _{d}d_{\sigma }^{\dag }d_{\sigma }\right)$, the
Hamiltonian (\ref{Eq1}) becomes
\begin{eqnarray}
H_{\mathrm{bom}} &=&\sum_{\sigma =\mathrm{cw,ccw}}g_{b}\left( d_{\sigma
}a_{\sigma }^{\dag }c_{\sigma }+d_{\sigma }^{\dag }a_{\sigma }c_{\sigma
}^{\dag }\right)  \notag \\
&&+\left( \Omega d_{\mathrm{cw}}^{\dag }+\Omega ^{\ast }d_{\mathrm{cw}%
}\right) .
\end{eqnarray}%
For a very strong control laser pumping to mode $d_{\mathrm{cw}}$, under the
condition that the power of the probe laser and the amplitude of acoustic
wave are very weak, we can treat the operator of the mode $d_{\mathrm{cw}}$
as a complex number as $\left\langle d_{\mathrm{cw}}\right\rangle =-i\Omega
/\kappa _{d}$, where $\kappa _{d}$ is the coupling strength between the
optical mode and fibres. Then, we obtain the linearized photon-phonon
interaction as
\begin{equation}
H_{\mathrm{bom}}\approx ga_{\mathrm{cw}}^{\dag }c_{\mathrm{cw}}+g^{\ast }a_{%
\mathrm{cw}}c_{\mathrm{cw}}^{\dag },
\end{equation}%
where $g\equiv g_{b}\left\langle d_{\mathrm{cw}}\right\rangle $. Without
loss of generality, we take $g$ as a positive-real number in the following.
We note that this strong driving enhanced beam-splitter-type photon-phonon
interaction has also been realized in the optomechanical systems for the WGM
optical modes coupling with the breathing mechanical mode~\cite{Dong2016NaPho,Ruesink2016NatCo,Shen2018NatCo,Ruesink2018NatCo}. So the results
in the following also applicable to the other WGM optomechanical systems.

The transmission spectra of the probe laser input from different port can
be obtained analytically by means of Fourier transformation method. The quantum Langevin equations (QLEs) for the operators are given
by
\begin{equation}
\frac{d}{dt}a_{\mathrm{cw}}=-\kappa _{a}a_{\mathrm{cw}}-igc_{\mathrm{cw}}+%
\sqrt{\kappa _{a}}a_{\mathrm{1,in}}+\sqrt{\kappa _{a}}a_{\mathrm{4,in}},
\end{equation}%
\begin{equation}
\frac{d}{dt}c_{\mathrm{cw}}=-\kappa _{c}c_{\mathrm{cw}}-iga_{\mathrm{cw}}+%
\sqrt{2\kappa _{c}}c_{\mathrm{cw,in}},
\end{equation}%
\begin{equation}
\frac{d}{dt}a_{\mathrm{ccw}}=-\kappa _{a}a_{\mathrm{ccw}}+\sqrt{\kappa _{a}}%
a_{\mathrm{2,in}}+\sqrt{\kappa _{a}}a_{\mathrm{3,in}},
\end{equation}%
where the optical modes $a_{\mathrm{cw}}$ and $a_{\mathrm{ccw}}$ are coupled
to both of the fibres with strength $\kappa _{a}$, and $a_{j,\mathrm{in}}$
is the field input from Port $j$; $\kappa _{c}$ is the acoustic damping rate
and $c_{\mathrm{cw,in}}$ is the field input into the acoustic mode. The
QLEs can be solved in the frequency domain by introducing the Fourier transform for an
operator $o$ as
\begin{equation}
\widetilde{o}\left( \omega \right) =\frac{1}{\sqrt{2\pi }}\int_{-\infty
}^{+\infty }o\left( t\right) e^{i\omega t}dt.
\end{equation}%
Based on the standard input-output theory~\cite{Gardiner1985PRA}, we get the expressions of
the output fields as
\begin{equation}
\widetilde{a}_{\mathrm{1,out}}\left( \omega \right) =S_{12}\left( \omega
\right) \widetilde{a}_{\mathrm{2,in}}\left( \omega \right) +S_{13}\left(
\omega \right) \widetilde{a}_{\mathrm{3,in}}\left( \omega \right) ,
\end{equation}%
\begin{eqnarray}
\widetilde{a}_{\mathrm{2,out}}\left( \omega \right) &=&S_{21}\left( \omega
\right) \widetilde{a}_{\mathrm{1,in}}\left( \omega \right) +S_{24}\left(
\omega \right) \widetilde{a}_{\mathrm{4,in}}\left( \omega \right)  \notag \\
&&+S_{2c}\left( \omega \right) \widetilde{c}_{\mathrm{cw,in}}\left( \omega
\right) ,
\end{eqnarray}%
\begin{eqnarray}
\widetilde{a}_{\mathrm{3,out}}\left( \omega \right) &=&S_{31}\left( \omega
\right) \widetilde{a}_{\mathrm{1,in}}\left( \omega \right) +S_{34}\left(
\omega \right) \widetilde{a}_{\mathrm{4,in}}\left( \omega \right)  \notag \\
&&+S_{3c}\left( \omega \right) \widetilde{c}_{\mathrm{cw,in}}\left( \omega
\right) ,
\end{eqnarray}%
where%
\begin{equation}
S_{12}\left( \omega \right) =\frac{\kappa _{a}}{\kappa _{a}-i\omega }-1,
\end{equation}%
\begin{equation}
S_{21}\left( \omega \right) =S_{34}\left( \omega \right) =\frac{\left(
\kappa _{c}-i\omega \right) \kappa _{a}}{\left( \kappa _{a}-i\omega \right)
\left( \kappa _{c}-i\omega \right) +g^{2}}-1,
\end{equation}%
\begin{equation}\label{SC1}
S_{13}\left( \omega \right) =\frac{\kappa _{a}}{\kappa _{a}-i\omega },
\end{equation}%
\begin{equation}\label{SC2}
S_{31}\left( \omega \right) =S_{24}\left( \omega \right) =\frac{\left(
\kappa _{c}-i\omega \right) \kappa _{a}}{\left( \kappa _{a}-i\omega \right)
\left( \kappa _{c}-i\omega \right) +g^{2}},
\end{equation}%
\begin{equation}
S_{2c}\left( \omega \right) =S_{3c}\left( \omega \right) =\frac{-ig\sqrt{%
2\kappa _{a}\kappa _{c}}}{\left( \kappa _{a}-i\omega \right) \left( \kappa
_{c}-i\omega \right) +g^{2}},
\end{equation}%
are the scattering coefficients. The transmission spectra are defined by
\begin{equation}
T_{ij}=\left\vert S_{ij}\left( \omega \right) \right\vert ^{2}
\end{equation}
for photons transport from Port $j$ to $i$.

The transmission spectra and the corresponding isolation between Port 1 and
2 are shown in Figs.~\ref{fig1}(b) and \ref{fig1}(c). We can see that the
photons transport unidirectionally from Port 1 to 2 around the frequency $%
\omega=0$, or from Port 2 to 1 around the frequencies $\omega=\pm g$. It is
worth emphasizing that the bandwidth for nonreciprocity with high isolation
is very narrow. We note that the bandwidth for nonreciprocity depends on the
optomechanical coupling strength $g$. To clarify this point further, we show
the bandwidth for $20$dB isolation as a function of $g$ in Fig.~\ref{fig1}%
(d). It shows that the bandwidth for $20$dB isolation increases with $g$ in
the weak coupling regime ($g<\kappa _{a}$), and then reaches the maximal
value about $\kappa _{a}/5$ in the strong coupling regime ($g>\kappa _{a}$).
The bandwidth with high isolation is one of the most important parameters for
nonreciprocal devices in practical applications. How to increase the
bandwidth with high isolation still needs more research.

Optical nonreciprocity can also be realized between the Port 1 and 3.
According to the transmission spectra and the corresponding isolation shown
in Figs.~\ref{fig1}(e) and \ref{fig1}(f), the photons transport
unidirectionally from Port 3 to 1 around the frequency $\omega=0$, or from
Port 1 to 3 around the frequencies $\omega=\pm g$. As shown in Fig.~\ref%
{fig1}(g), the bandwidth for $-20$dB isolation increases monotonously with
coupling strength $g$, and most importantly, the value of bandwidth is not
saturated in the strong coupling regime. So we can obtain a much broader
bandwidth for high isolation between the Port 1 and 3. How to get a broader
bandwidth with a higher isolation is the main issues discussed in this
paper. We will show that both the bandwidth and isolation for
optical nonreciprocity can be improved in a 1D optomechanical array via nonreciprocal band
structure.

\section{Nonreciprocal Band Structure}\label{NBS}

\begin{figure}[tbp]
\includegraphics[bb=63 75 513 791, width=8.5 cm, clip]{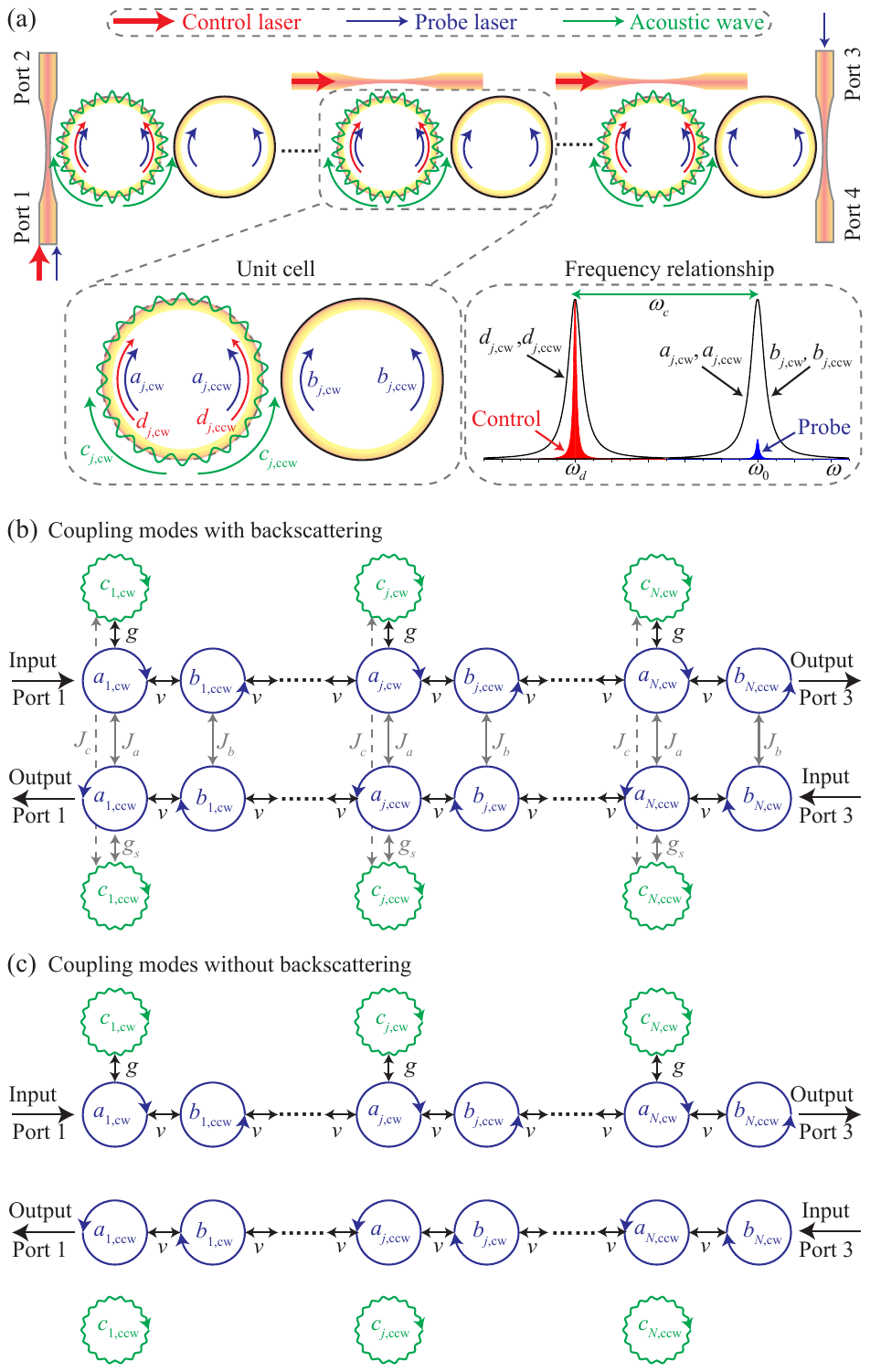}
\caption{(Color online) (a) Schematic of a 1D Brillouin optomechanical array
containing $N$ unit cells and the input-output waveguides. (b) Geometric
structure of the lattice with the backscattering effect taken account $J_{\eta}\neq 0$. (c) Geometric structure of the lattice without taking account of the backscattering effect $J_{\eta} =0$.}
\label{fig2}
\end{figure}

We propose a 1D optomechanical array with $N$ unit cells as shown in Fig.~\ref{fig2}(a), where
the unit cell is consisting of a Brillouin optomechanical system coupled to
a WGM microresonator. In the rotating frame defined by the unitary
transformation operator $R_N\left( t\right)= \exp (-iH_N t)$ with $%
H_N=\sum_{\sigma =\mathrm{cw,ccw}}\sum_{j =1}^{N}\left( \omega _{0}a_{\sigma
}^{\dag }a_{\sigma }+\omega _{0}b_{\sigma }^{\dag }b_{\sigma }+\omega
_{c}c_{\sigma }^{\dag }c_{\sigma }+\omega _{d}d_{\sigma }^{\dag }d_{\sigma
}\right)$, the system can be described by the total Hamiltonian
\begin{equation}
H_{\mathrm{tot}}=H_{\mathrm{1\rightarrow 3}}^{\left( 0\right) }+H_{\mathrm{%
3\rightarrow 1}}^{\left( 0\right) }+H_{\mathrm{BS}}^{\left( 0\right) },
\end{equation}%
where $H_{\mathrm{1\rightarrow 3}}^{\left( 0\right) }$ is the Hamiltonian
for photons transport from Port 1 to 3,
\begin{eqnarray}
H_{\mathrm{1\rightarrow 3}}^{\left( 0\right) } &=&\sum_{j=1}^{N}\left(
g_{j}d_{j,\mathrm{cw}}^{\dag }a_{j,\mathrm{cw}}c_{j,\mathrm{cw}}^{\dag
}+\Omega _{j}d_{j,\mathrm{cw}}^{\dag }+va_{j,\mathrm{cw}}b_{j,\mathrm{ccw}%
}^{\dag }\right)  \notag \\
&&+\sum_{j=1}^{N-1}va_{j+1,\mathrm{cw}}b_{j,\mathrm{ccw}}^{\dag }+\mathrm{%
h.c.,}
\end{eqnarray}%
$H_{\mathrm{3\rightarrow 1}}^{\left( 0\right) }$ is the one for photons
transport in the reverses direction, i.e., from Port 3 to 1,
\begin{eqnarray}
H_{\mathrm{3\rightarrow 1}}^{\left( 0\right) } &=&\sum_{j=1}^{N}\left(
g_{j}d_{j,\mathrm{ccw}}^{\dag }a_{j,\mathrm{ccw}}c_{j,\mathrm{ccw}}^{\dag
}+va_{j,\mathrm{ccw}}b_{j,\mathrm{cw}}^{\dag }\right)  \notag \\
&&+\sum_{j=1}^{N-1}va_{j+1,\mathrm{ccw}}b_{j,\mathrm{cw}}^{\dag }+\mathrm{%
h.c.,}
\end{eqnarray}%
and
\begin{equation}
H_{\mathrm{BS}}^{\left( 0\right) }=\sum_{j=1}^{N}\sum_{\eta =a,b,c,d}J_{\eta
}\left( \eta _{j,\mathrm{cw}}^{\dag }\eta _{j,\mathrm{ccw}}+\eta _{j,\mathrm{%
ccw}}^{\dag }\eta _{j,\mathrm{cw}}\right),
\end{equation}%
is the backscattering induced interaction term for the photons transport in
different directions.

To realize nonreciprocal band structure, a strong control laser is pumped to
the mode $d_{j,\mathrm{cw}}$. Under the conditions that the power of the
probe laser and the amplitude of acoustic wave are very weak, the modes $%
d_{j,\mathrm{cw}}$ and $d_{j,\mathrm{ccw}}$ can be treated classically as
complex numbers as
\begin{equation}
\left\langle d_{j,\mathrm{cw}}\right\rangle =\frac{-i2\Omega _{j}\kappa _{d}%
}{\kappa _{d}^{2}+4J_{d}^{2}}
\end{equation}%
and
\begin{equation}
\left\langle d_{j,\mathrm{ccw}}\right\rangle =\frac{-i2J_{d}}{\kappa _{d}}%
\left\langle d_{j,\mathrm{cw}}\right\rangle.
\end{equation}%
So we obtain the linearized Hamiltonian
\begin{equation}  \label{Hlin}
H_{\mathrm{lin}}=H_{\mathrm{1\rightarrow 3}}+H_{\mathrm{3\rightarrow 1}}+H_{%
\mathrm{BS}},
\end{equation}%
where%
\begin{eqnarray}  \label{H31}
H_{\mathrm{1\rightarrow 3}} &=&\sum_{j=1}^{N}\left( ga_{j,\mathrm{cw}}c_{j,%
\mathrm{cw}}^{\dag }+va_{j,\mathrm{cw}}b_{j,\mathrm{ccw}}^{\dag }\right)
\notag \\
&&+\sum_{j=1}^{N-1}va_{j+1,\mathrm{cw}}b_{j,\mathrm{ccw}}^{\dag }+\mathrm{%
h.c.,}
\end{eqnarray}%
\begin{eqnarray}  \label{H13}
H_{\mathrm{3\rightarrow 1}} &=&\sum_{j=1}^{N}\left( g_{s}a_{j,\mathrm{ccw}%
}c_{j,\mathrm{ccw}}^{\dag }+va_{j,\mathrm{ccw}}b_{j,\mathrm{cw}}^{\dag
}\right)  \notag \\
&&+\sum_{j=1}^{N-1}va_{j+1,\mathrm{ccw}}b_{j,\mathrm{cw}}^{\dag }+\mathrm{%
h.c.,}
\end{eqnarray}%
and
\begin{equation}
H_{\mathrm{BS}}=\sum_{j=1}^{N}\sum_{\eta =a,b,c}J_{\eta }\left( \eta _{j,%
\mathrm{cw}}^{\dag }\eta _{j,\mathrm{ccw}}+\eta _{j,\mathrm{ccw}}^{\dag
}\eta _{j,\mathrm{cw}}\right),
\end{equation}
where $g\equiv g_{j}\left\langle d_{j,\mathrm{cw}}\right\rangle $ and $%
g_{s}\equiv g_{j}\left\langle d_{j,\mathrm{ccw}}\right\rangle$ are the
pumping enhanced photon-phonon coupling strengths; see Fig.~\ref{fig2}(b).
For simplicity, we set $g$ as a positive-real number.
We should point out that $g_{s}$ is induced by the backscattering as $g_{s}
=(-i2J_{d}/\kappa_{d})g$, and the differences between $g$ and $g_{s}$ is the
key ingredient for the nonreciprocal band structure.

\begin{figure}[tbp]
\includegraphics[bb=52 295 522 624, width=8.5 cm, clip]{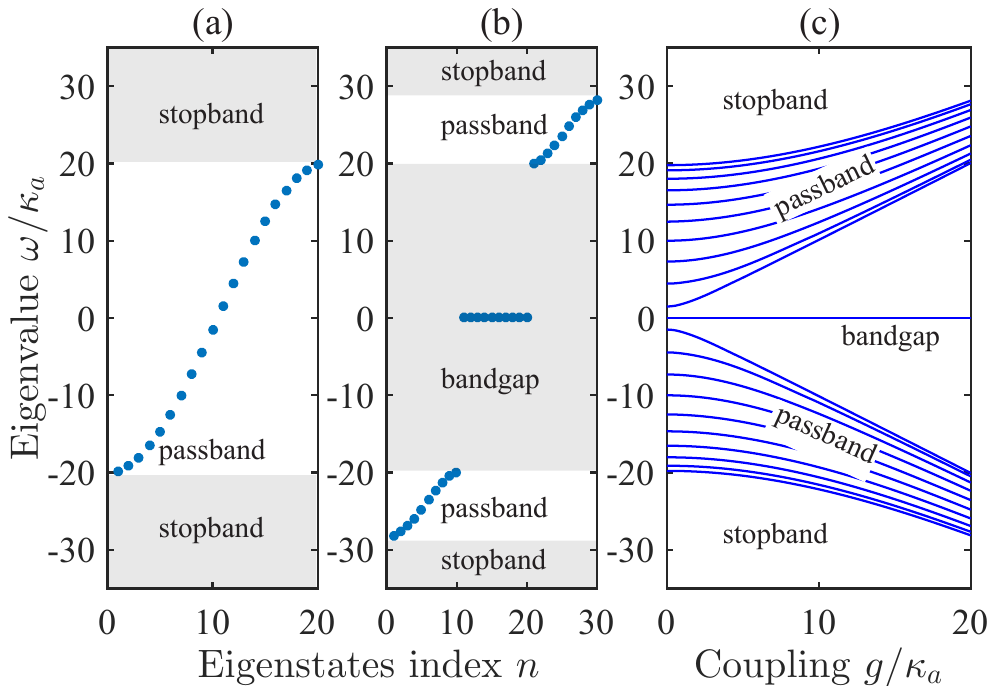}
\caption{(Color online) Band structure of (a) a SSH lattice for
the photons transport from Port 3 to 1 and (b) a stub
lattice for the photons transport from Port 1
to 3. (c) Band structures of the stub lattice versus the coupling
strength $g/\kappa_a$. The other parameters are $N=10$, $v=10\kappa_a$ and $g=20\kappa_a$ in (b).}
\label{fig3}
\end{figure}

Let us analyze the band structure and the corresponding
transmission spectra by neglecting the backscattering, i.e., $g_s=0$ and $%
J_{\eta}=0$ first. In this case, the geometric structure of the coupled modes are
divided into three parts [see Fig.~\ref{fig2}(c)]: (i) a stub lattice~\cite{Hyrkas2013PRA,Baboux2016PRL,Gabriel2022PRR} for
the photons transport from Port 1 to 3, (ii) a Su-Schrieffer-Heeger (SSH)
lattice~\cite{SSH1979PRL,Janos2016Springer} for the photons transport from Port 3 to 1, and (iii) the isolated
acoustic modes $c_{j,\mathrm{ccw}}$. The Hamiltonian of the stub lattice is
given by $H_{\mathrm{1\rightarrow 3}}$ (\ref{H31}), and the SSH lattice and
the isolated acoustic modes are described by $H_{\mathrm{3\rightarrow 1}}$ (%
\ref{H13}) with $g_s=0$. The band structure of the SSH and stub lattices for photons
transport in different direction can be found by numerically solving the
eigenvalues of Eqs.~(\ref{H31}) and (\ref{H13}), respectively. The band
structures for the lattices containing $N$ unit cells ($N=10$) are shown
in Fig.~\ref{fig3}(a) and \ref{fig3}(b), respectively. There is only one
passband in the band structures of the SSH lattice for photons transport
from Port 3 to 1. In contrast, there are two passbands in the band
structures of the stub lattice for the photons transport from Port 1 to 3,
separated by a bandgap induced by the photon-phonon interaction $g$. The
width of the bandgap becomes broader with the increasing of $g$, as shown in Fig.~\ref{fig3}(c).

The band structures can also be analyzed analytically in the momentum space under the periodic
boundary condition. By
introducing the Fourier transformation $O_k=(1/\sqrt{N}%
)\sum_{j}e^{ijkd_{0}}O_{j}$ ($k$ is the wave number and $d_0$ is the lattice
constant, hereafter we set $d_0=1$ for simplicity), the Hamiltonian (\ref{Hlin}) can be rewritten as
\begin{equation}
H_{\mathrm{lin}}=\sum_{k}V_{k}^{\dag }H_{\mathrm{lin}}\left( k\right) V_{k},
\end{equation}%
where $V_{k}^{\dag }=( a_{k,\mathrm{cw}}^{\dag }, b_{k,\mathrm{ccw}}^{\dag
}, c_{k,\mathrm{cw}}^{\dag }, a_{k,\mathrm{ccw}}^{\dag }, b_{k,\mathrm{cw}%
}^{\dag }, c_{k,\mathrm{ccw}}^{\dag }) $, and the Hamiltonian in the momentum space is given by
\begin{equation}
H_{\mathrm{lin}}\left( k\right) =\left(
\begin{array}{cc}
H_{\mathrm{1\rightarrow 3}}\left( k\right) & H_{\mathrm{BS}}\left( k\right)
\\
H_{\mathrm{BS}}\left( k\right) & H_{\mathrm{3\rightarrow 1}}\left( k\right)%
\end{array}%
\right).
\end{equation}%
with the submatrices
\begin{equation}  \label{Hk_stub}
H_{\mathrm{1\rightarrow 3}}\left( k\right) =\left(
\begin{array}{ccc}
0 & \rho & g \\
\rho^{\ast} & 0 & 0 \\
g & 0 & 0%
\end{array}%
\right) ,
\end{equation}%
\begin{equation}  \label{Hk_ssh}
H_{\mathrm{3\rightarrow 1}}\left( k\right) =\left(
\begin{array}{ccc}
0 & \rho & g_s \\
\rho^{\ast} & 0 & 0 \\
g_s^{\ast} & 0 & 0%
\end{array}%
\right) ,
\end{equation}%
\begin{equation}
H_{\mathrm{BS}}\left( k\right) =\left(
\begin{array}{ccc}
J_{a} & 0 & 0 \\
0 & J_{b} & 0 \\
0 & 0 & J_{c}%
\end{array}%
\right).
\end{equation}
Here we define $\rho\equiv v+ve^{ik}$.

The frequency spectrum of the lattices can be read off from Eqs.~(\ref%
{Hk_stub}) and (\ref{Hk_ssh}) for $g_{s}=0$ and $J_{\eta }=0$. The
eigenvalues of Eqs.~(\ref{Hk_stub}) and (\ref{Hk_ssh}) can be written in an
unified form as
\begin{equation}
\omega (k)=\left\{
\begin{array}{c}
\sqrt{|\rho |^{2}+g_{\rm om}^{2}}, \\
0, \\
-\sqrt{|\rho |^{2}+g_{\rm om}^{2}},%
\end{array}%
\right.   \label{Ek}
\end{equation}%
where $g_{\rm om}=0$ for the SSH lattice and $g_{\rm om}= g$ for the stub lattice. As shown in Fig.~\ref%
{fig4}(a), there is only one passband from $-2v$ to $2v$ (width $4v$) in the band structure for
the photons transport from Port 3 to 1 (SSH lattice), where the eigenvalues $\omega (k)=0$ for the isolated acoustic modes are not shown here. In contrast, there is a bandgap (from $g$ to $-g$) between the
two passbands (from $g$ to $\sqrt{4v^{2}+g^{2}}$ and from $-g$ to $-\sqrt{%
4v^{2}+g^{2}}$) in the band structure of stub lattice for the photons transport from Port 1
to 3 [see Fig.~\ref{fig4}(b)].

\section{Broadband Optical Nonreciprocity}\label{WBON}

Now, we discuss the transmission spectra between the Ports 1 and 3 based on nonreciprocal band structure.
After introducing the decay terms and the corresponding input fields, the QLEs for the operators are given by
\begin{eqnarray}
\frac{d}{dt}a_{j,\mathrm{cw}} &=&-ivb_{j,\mathrm{ccw}}-i v b_{j-1,\mathrm{ccw}%
}-igc_{j,\mathrm{cw}}-iJ_{a}a_{j,\mathrm{ccw}}  \notag \\
&&-\frac{\kappa _{a}}{2}a_{j,\mathrm{cw}}+\sqrt{\kappa _{a}}a_{j,\mathrm{%
cw,in}},
\end{eqnarray}%
\begin{eqnarray}
\frac{d}{dt}b_{j,\mathrm{ccw}} &=&-iva_{j,\mathrm{cw}}-i v a_{j+1,\mathrm{cw}%
}-iJ_{b}b_{j,\mathrm{cw}}  \notag \\
&&-\frac{\kappa _{b}}{2}b_{j,\mathrm{ccw}}+\sqrt{\kappa _{b}}b_{j,\mathrm{%
ccw,in}},
\end{eqnarray}%
\begin{eqnarray}
\frac{d}{dt}c_{j,\mathrm{cw}} &=&-iga_{j,\mathrm{cw}}-iJ_{c}c_{j,\mathrm{ccw}%
}  \notag \\
&&-\frac{\kappa _{c}}{2}c_{j,\mathrm{cw}}+\sqrt{\kappa _{c}}c_{j,\mathrm{%
cw,in}},
\end{eqnarray}%
\begin{eqnarray}
\frac{d}{dt}a_{j,\mathrm{ccw}} &=&-ivb_{j,\mathrm{cw}}-i v b_{j-1,\mathrm{cw}%
}-ig_{s}c_{j,\mathrm{ccw}}-iJ_{a}a_{j,\mathrm{cw}}  \notag \\
&&-\frac{\kappa _{a}}{2}a_{j,\mathrm{ccw}}+\sqrt{\kappa _{a}}a_{j,\mathrm{%
ccw,in}},
\end{eqnarray}%
\begin{eqnarray}
\frac{d}{dt}b_{j,\mathrm{cw}} &=&-iva_{j,\mathrm{ccw}}-iv a_{j+1,\mathrm{ccw}%
}-iJ_{b}b_{j,\mathrm{ccw}}  \notag \\
&&-\frac{\kappa _{b}}{2}b_{j,\mathrm{cw}}+\sqrt{\kappa _{b}}b_{j,\mathrm{%
cw,in}},
\end{eqnarray}%
\begin{eqnarray}
\frac{d}{dt}c_{j,\mathrm{ccw}} &=&-ig^{\ast}_{s}a_{j,\mathrm{ccw}}-iJ_{c}c_{j,%
\mathrm{cw}}  \notag \\
&&-\frac{\kappa _{c}}{2}c_{j,\mathrm{ccw}}+\sqrt{\kappa _{c}}c_{j,\mathrm{%
ccw,in}},
\end{eqnarray}%
where $\kappa_\eta$ ($\eta=a,\: b,\: c$) are the decay rate of the optical and mechanical modes, and $ \eta_{j,\mathrm{\sigma,in}}$ ($\sigma= {\rm cw,\: ccw}$) are the input operators of these
modes.
For the sake of brevity, we rewrite the QLEs in a matrix form as
\begin{equation}
\frac{d}{dt}V=-MV+\sqrt{\Gamma }V_{\mathrm{in}},
\end{equation}%
where $\left( V\right) ^{T}=\left( \left( V_{\mathrm{1\rightarrow 3}}\right)
^{T},\left( V_{\mathrm{3\rightarrow 1}}\right) ^{T}\right) $, $\left( V_{%
\mathrm{1\rightarrow 3}}\right) ^{T}=\left( \cdots ,a_{j,\mathrm{cw}},b_{j,%
\mathrm{ccw}},c_{j,\mathrm{cw}},\cdots \right) $ , $\left( V_{\mathrm{%
3\rightarrow 1}}\right) ^{T}=\left( \cdots ,a_{j,\mathrm{ccw}},b_{j,\mathrm{%
cw}},c_{j,\mathrm{ccw}},\cdots \right) $, $\left( V_{\mathrm{in}}\right)
^{T}=\left( \left( V_{\mathrm{1\rightarrow 3,in}}\right) ^{T},\left( V_{%
\mathrm{3\rightarrow 1,in}}\right) ^{T}\right) $, $\left( V_{\mathrm{%
1\rightarrow 3,in}}\right) ^{T}=\left( \cdots ,a_{j,\mathrm{cw,in}},b_{j,%
\mathrm{ccw,in}},c_{j,\mathrm{cw,in}},\cdots \right) $, $\left( V_{\mathrm{%
3\rightarrow 1,in}}\right) ^{T}=\left( \cdots ,a_{1,\mathrm{ccw,in}},b_{1,%
\mathrm{cw,in}},c_{1,\mathrm{ccw,in}},\cdots \right) $, $\Gamma =\mathrm{diag%
}\left( \cdots ,\kappa _{a},\kappa _{b},\kappa _{c},\cdots \right) $, and $M$
is a $6N\times 6N$ coefficient matrix.

We solve the QLEs in the frequency domain and get the expression
\begin{equation}
\widetilde{V}\left( \omega \right) =\left( M-i\omega I\right) ^{-1}\sqrt{%
\Gamma }\widetilde{V}_{\mathrm{in}}\left( \omega \right),
\end{equation}%
where $I$ is the identity matrix. Based on the input-output theory~\cite{Gardiner1985PRA}, the output vector
$\left( V_{\mathrm{out}}\right) ^{T}=\left( \left( V_{\mathrm{%
1\rightarrow 3,out}}\right) ^{T},\left( V_{\mathrm{3\rightarrow 1,out}%
}\right) ^{T}\right) $, $\left( V_{\mathrm{1\rightarrow 3,out}}\right)
^{T}=\left( \cdots ,a_{j,\mathrm{cw,out}},b_{j,\mathrm{ccw,out}},c_{j,%
\mathrm{cw,out}},\cdots \right) $, $\left( V_{\mathrm{3\rightarrow 1,out}%
}\right) ^{T}=\left( \cdots ,a_{1,\mathrm{ccw,out}},b_{1,\mathrm{cw,out}%
},c_{1,\mathrm{ccw,out}},\cdots \right) $, in the frequency domain is obtained as
\begin{equation}
\widetilde{V}_{\mathrm{out}}\left( \omega \right) =U\left( \omega \right)
\widetilde{V}_{\mathrm{in}}\left( \omega \right),
\end{equation}%
where
\begin{equation}\label{Uw}
U\left( \omega \right) =\sqrt{\Gamma }\left( M-i\omega I\right) ^{-1}\sqrt{%
\Gamma }-I.
\end{equation}%
The transmission spectrum for the photons transport from port 1 to 3 is
given by
\begin{equation}
T_{31}\left( \omega \right) =\left\vert U_{\left( 3N-1\right) ,1}\left(
\omega \right) \right\vert ^{2},
\end{equation}%
and the transmission spectrum in the reverse direction is given by
\begin{equation}
T_{13}\left( \omega \right) =\left\vert U_{\left( 3N+1\right) ,\left(
6N-1\right) }\left( \omega \right) \right\vert ^{2},
\end{equation}
where $U_{ij}(\omega)$ is the element at the $i$th row and $j$th column of the scattering matrix $U(\omega)$ in Eq.~(\ref{Uw}).

\begin{figure*}[tbp]
\includegraphics[bb=16 301 581 587, width=16 cm, clip]{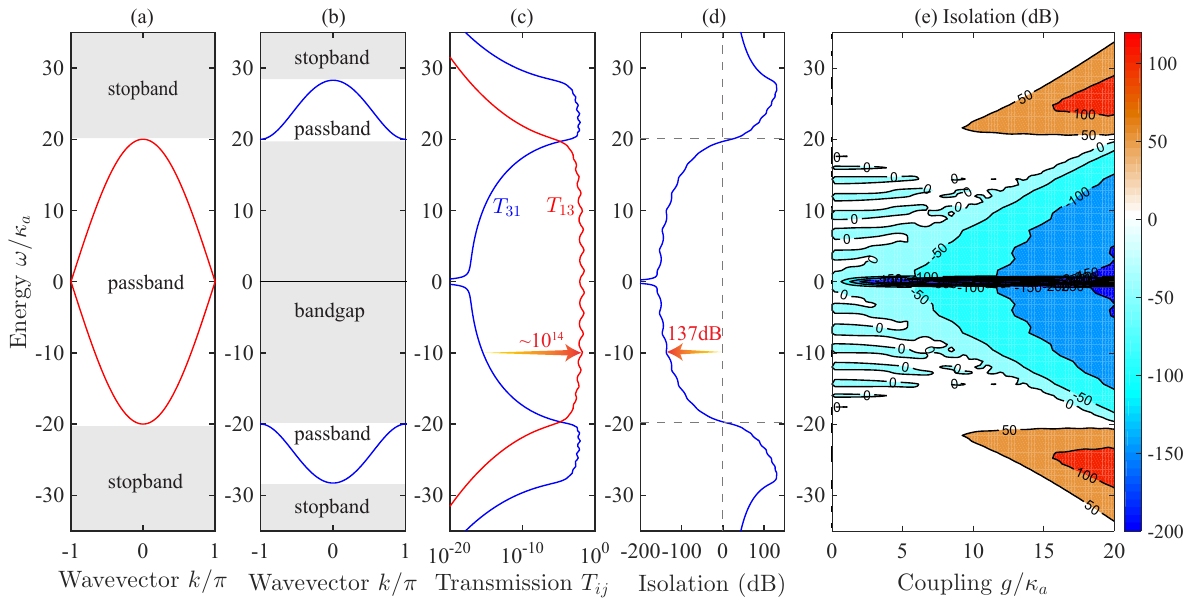}
\caption{(Color online) Frequency spectrum for photons transport (a) from Port 3 to 1 and (b) from Port 1 to 3. (c) The transmission spectra ($T_{13}$ and $T_{31}$) and (d) the isolation $I=10\log_{10}(T_{31}/T_{13})$ versus the energy of the input photons for $N=10$. The other parameters are $g=2v$, $v=10\kappa_a$, $\kappa_a=\kappa_b$,
$\protect\kappa_c=\kappa_a/100$, and $J_{\protect\eta}=0$ ($\protect\eta=a,\,b,\,c,\,d$).}
\label{fig4}
\end{figure*}

We note that there is a transmission window in the transmission spectrum for the passband, and the photon transport is suppressed significantly in the stopband or bandgap.
To obtain great optical nonreciprocity, we set $g=2v$, so the passband ($-2v< \omega < 2v$) for photons transport from Port 3 to 1 corresponds to the bandgap ($-g<\omega<g$) for photons transport from Port 1 to 3 with two passbands ($-\sqrt{4v^2+g^2}<\omega<-g$ and $g<\omega<\sqrt{4v^2+g^2}$), as shown in Figs.~\ref{fig4}(a) and \ref{fig4}(b).
In this case, the transmission spectra ($T_{31}$ and $T_{13}$) and the corresponding isolation ($I=10\log_{10}(T_{31}/T_{13})$) are shown in Figs.~\ref{fig4}(c) and \ref{fig4}(d).
We obtain strong nonreciprocity ($\sim 140$dB) with a broad bandwidth ($\sim g\gg \kappa_a$) based on the nonreciprocal band structure.
In addition, the width of the strong nonreciprocity can be tuned by the coupling strength $g$ [see Fig.~\ref{fig4}(e)], which depends on the optical driving strength.

\begin{figure}[tbp]
\includegraphics[bb=110 210 441 590, width=7 cm, clip]{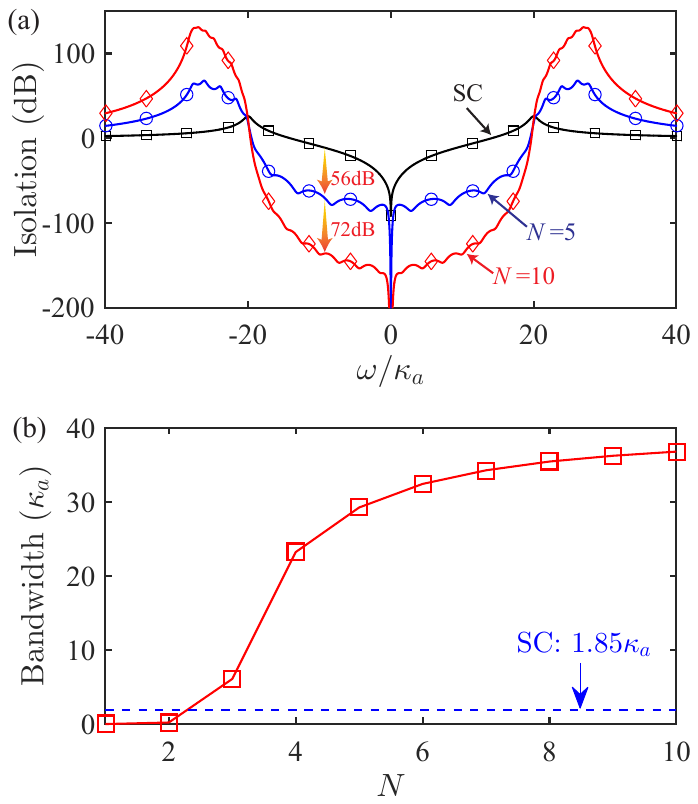}
\caption{(Color online) (a) The isolation $10\log_{10}(T_{31}/T_{13})$ as a function of energy $\omega/\kappa_a$. (b) The
bandwidth with isolation of $-50$dB as a function of the number of unit cells $N$. The dashed line is the bandwidth for optomechanical nonreciprocity in a single cavity. The other
parameters are $v=10\kappa_a$, $g=2v$, $\kappa_b=\kappa_a$, $\kappa_c=\kappa_a/100$, and $J_{\protect\eta}=0$ ($\protect\eta=a,\,b,\,c,\,d$).}
\label{fig5}
\end{figure}

\begin{figure*}[tbp]
\includegraphics[bb=33 331 573 617, width=16 cm, clip]{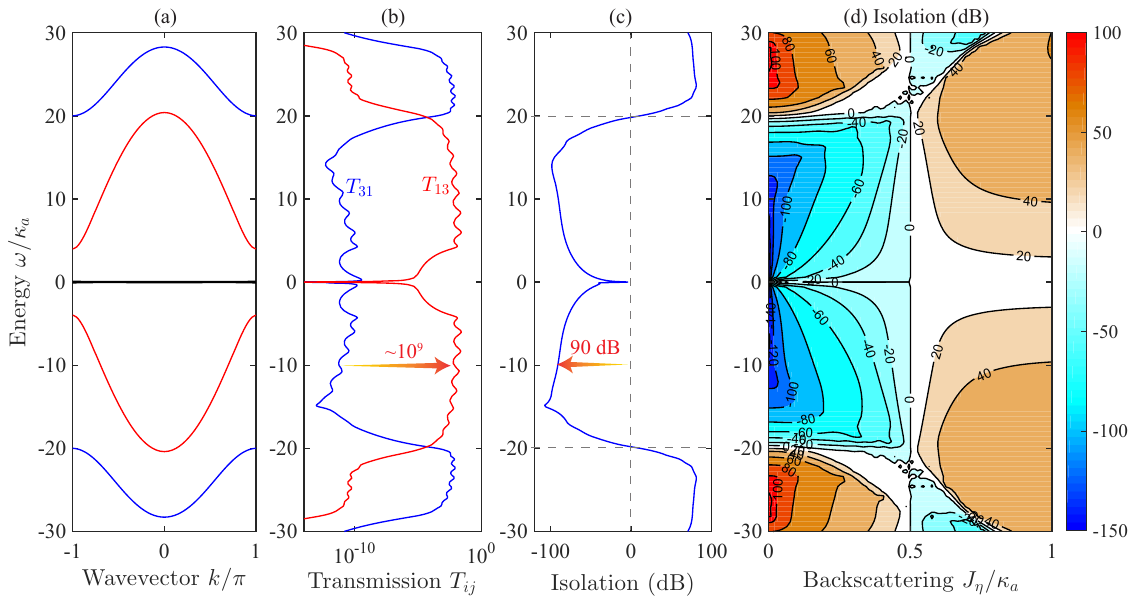}
\caption{(Color online) (a) Frequency spectrum of the Hamiltonian (\ref{Hlin}) as a function of the wavevector $k/\pi$, (b) the
transmission spectra ($T_{13}$ and $T_{31}$) and (c) the isolation $10\log_{10}(T_{31}/T_{13})$ versus energy $\omega/\kappa_a$, for $J_{\eta}=0.1\kappa_a$ ($\protect\eta=a,\,b,\,c,\,d$). (d) The isolation obtained as a function of energy $\omega/\kappa_a$ and backscattering $J_{\eta}/\kappa_a$. The other parameters are $v=10\kappa_a$, $g=2v$, $\kappa_b=\kappa_d=\kappa_a$, $\kappa_c=\kappa_a/100$, and $N=10$.}
\label{fig6}
\end{figure*}

To show the advantageous of optical nonreciprocity via the nonreciprocal band structure to the nonreciprocity in the Brillouin optomechanical system with single cavity [see Eqs.~(\ref{SC1}) and (\ref{SC2})], we show the isolations for different systems in Fig.~\ref{fig5}(a) [SC stands for single cavity].
In comparison to nonreciprocity in a single cavity, both the isolation and bandwidth are dramatically improved for the nonreciprocity in an optomechanical array with nonreciprocal band structure.
Specifically, the isolation is improved by 56dB for an optomechanical array with $N=5$ unit cells, and it can be improved further by $72$dB when the unit cells increases to $N=10$.
Moreover, we show the band width with $-50$dB isolation versus the number of unit cells $N$ in Fig.~\ref{fig5}(b).
Clearly, the advantageous of the nonreciprocity in an optomechanical array starts to appear with the number of unit cells $N=3$, and the bandwidth gradually tends toward $4v$ with the increase of the number of unit cells.

Finally, let us discuss the influences of the backscattering on the nonreciprocity based on nonreciprocal band structure.
The backscattering effect induces the coupling between the path for photons transport from Port 1 to 3 and the path for photons transport from Port 3 to 1, as shown in Fig.~\ref{fig2}(b).
In this case, the frequency spectrum as a function of the wavevector $k/\pi$ is shown in Fig.~\ref{fig6}(a), which is the combination of the energy bands for photons transport in bidirection between Ports 1 and 3.
Besides, the backscattering effect induces a nonzero photon-phonon coupling $g_s \neq 0$ for the photons transport from Port 3 to 1, which led to the appearing of a bandgap $-|g_s|<\omega<|g_s|$ in the band structure.
When the backscattering effect is weak, the nonreciprocity still can be obtained with high isolation and broad bandwidth as shown in Figs.~\ref{fig6}(b) and \ref{fig6}(c) for $J_{\eta}=0.1\kappa_a$.
To show the effect of backscattering on the nonreciprocity clearly, the isolation as a function of energy $\omega/\kappa_a$ and backscattering $J_{\eta}/\kappa_a$ is shown in Fig.~\ref{fig2}(d).
We can see that the nonreciprocal effect becomes weaker with the increasing of $J_{\eta}$ in the weak backscattering regime ($J_{\eta}<\kappa_{a}/2$) and disappears when $J_{\eta}=\kappa_{a}/2$.
This can be understood from the relation $g_{s}=(-i2J_{d}/\kappa_{d})g$, which indicates that the difference between $g_s$ and $g$ becomes smaller with the increasing of $J_{\eta}$, and $|g_{s}|=g$ for $\kappa_{d}=\kappa_{a}$ and $J_{d}=\kappa_{d}/2$.
The permitting transport direction even changes when $J_{\eta}>\kappa_{a}/2$ for we have $|g_{s}|>g$ in the strong backscattering regime.

\section{Conclusions}\label{Con}

In conclusion, we have revealed the challenge in achieving nonreciprocal isolator with both broad bandwidth and high isolation in an optomechanical system with single cavity.
To overcome this challenge, we proposed to realize broadband optical nonreciprocity in a 1D optomechanical array with nonreciprocal band structure.
We investigated the nonreciprocal band structure in a 1D Brillouin optomechanical array with directional enhanced optomechanical interaction by directional optical pumping, and demonstrated optical nonreciprocity with both broad bandwidth and high isolation  in a controllable way.
Looking forwards, such optomechanical lattices offer a path to realize proposals exploring the nonreciprocal collective effects, such as nonreciprocal topological photonic and phononic phases~\cite{RenHJ2022NatCo,Youssefi2021arXiv}, and viewed more broadly, it can be used to explore exotic quantum light-matter interactions in nonreciprocal optomechanical lattices~\cite{DongXL2021PRL,DongXL2023PRB}.


\begin{acknowledgments}
This work is supported by the
National Natural Science Foundation of China (NSFC) (Grants No.~12064010 and  No.~12247105),
Natural Science Foundation of Hunan Province of China (Grant
No.~2021JJ20036), and the science and technology innovation Program of Hunan Province (Grant No.~2022RC1203).
\end{acknowledgments}

\bibliography{ref}


\end{document}